\documentclass{article} %使用中文版的article文档类型排版，并选择UTF8编码格式
\usepackage{amsmath}  %使用宏包，这里使用的是调用公式宏包，可以调用多个宏包
\usepackage{multicol}  %分栏
\usepackage{graphicx}   %我们先添加宏包插图，方便接下来添加图片
\usepackage{ragged2e}   %与justifying指令实现两端对齐
\usepackage{amsthm}
\usepackage{booktabs}
\usepackage{array}
\usepackage{float}   %用于双栏下插入图片
\usepackage{caption}     %图片脚注
\usepackage{subfigure}   %子图包
\usepackage{fancyhdr}
\usepackage{indentfirst} %添加作者信息
\usepackage{geometry}
\usepackage{cuted}
\usepackage{setspace}
\usepackage{listings}   %放置代码块
\usepackage{xcolor}
\usepackage{url}
\usepackage{chngcntr}
\counterwithin{figure}{section}
\counterwithin{table}{section}
\begin{document}  %开始写文章
\title{\LARGE\bf Move first, and become unbeatable: Strategy study of different Tic-tac-toe}	
\author{Junan Pan}
\date{}		
\maketitle     %我们写了以上内容以后一定要添加这个，制作标题，否则上面的内容都是无效的。

\begin{spacing}{1.5}
\large{\bf Abstract:} The main challenge of combinatorial game theory is to handle combinatorial chaos, if one player knows the strategy better than his opponent, he is able to determine the exact results of a game. If both players are qualified competitor, the result usually depends on the order when they take turns, which however, may not be determined by a player casually. 

So, this research studied different kinds of tic-tac-toe, assuming that a player always moves first, tried to figure out the winning or unbeatable strategy for the first player in different kinds of tic-tac-toe. It is titled “Move first, and become unbeatable”, which also reflects the philosophy in life.

 
 \section{Introduction}
\large{\bf}  Combinatorial Games are designed for two players playing alternate moving. Each player knows all the rules and information about the game. There is no randomization mechanism such as flipping a coin or rolling a die or some other hidden rules. Tic-tac-toe is a paper-and-pencil game for two players who take turns marking the spaces in a three-by-three grid with X or O. In the normal version, the player who succeeds in placing three of their marks in a horizontal, vertical, or diagonal row is the winner.

 \section{Unbeatable strategy study in different tic-tac-toe}

\subsection{Normal version} 
\large{\bf Rule Introduction}  
As described in introduction, the normal version of tic-tac-toe is a game a game in which two players alternately put X and O in compartments of a figure formed by two vertical lines crossing two horizontal lines and each tries to get a row of three X or three O before the opponent does. One of a typical winning result for O is shown in Figure 2.1.
\begin{figure}[H]
  \centering
  \includegraphics[width=4.5cm]{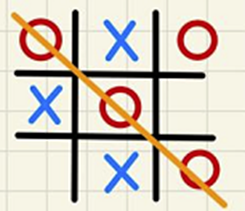}
  \caption{Red O wins in normal version}
\end{figure}
\large{\bf Strategy Analysis}  
 The normal version of tic-tac-toe is a solved game. It always ends up with draw if both players play their best, so for the first player, the minimum requirement for him is a draw. If it is possible, he should also try to win his unqualified opponent by adopting the method with the highest winning strategy.

Statistically speaking, the first player takes advantage not only because he has one more step to move, but also because he could occupy the optimal central position directly, which greatly improves his probability of making a row compared to other positions. It is also possible for the first player to win if he first occupies the corner of edge. However, he just needs to choose a place which keeps him more far away from losing the game.

The central position has been claimed now, what would the opponent do? There are only 2 situations for the center position: edge or corner.
\begin{figure}[H]
  \centering
  \includegraphics[width=10cm]{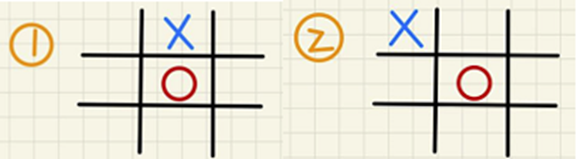}
  \caption{Two situations the first player would meet after occupying center}
\end{figure}
Actually, there is a very famous algorithm called minimax (MM) in Game Theory, which is used to minimize the possible loss for a worst case (maximum loss) scenario, and can be used in tic-tac-toe too. But this algorithm needs state search and calculation, including DFS, alpha-beta pruning and so on, which is not a very practical for a real player to operate.

The aim of this research is just finding a simple but effective unbeatable strategy for the first player, so nothing is better than reducing the uncertainty and let as much as possible things under his control from the beginning. After some attempts, this method does exist.

$\textcircled{1}$If the second player first occupies the edge. From now on it is N-position for the first player. (N-Positions that are winning for the next player, who is about to move from the current position.) The winning strategy is quite easy for the first player and the entire process is shown in Figure 2.3.
\begin{figure}[H]
  \centering
  \includegraphics[width=8cm]{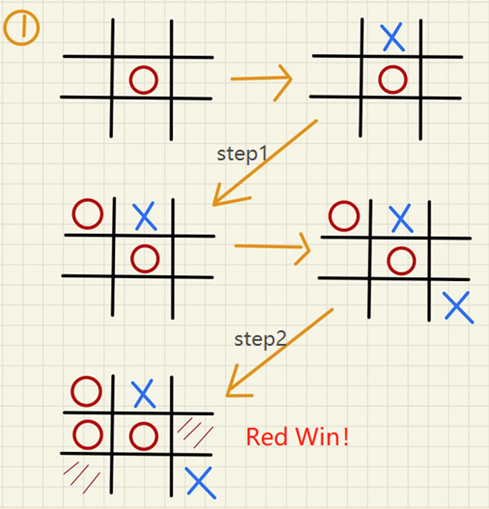}
  \caption{How the first player can force to win after his opponent occupied the edge}
\end{figure}
The first player puts his second chess at the top left corner, the second player is forced to occupy the bottom right corner to defend. Then the first player occupies the edge between his previous chesses, then he could win by claiming any shadowed position of the last chessboard in Figure 2.3, regardless where the blue X claims.

$\textcircled{2}$If the second player first occupies the corner.
When meeting a qualified opponent, the first player will always face this situation. The result starts tending to a draw. Figure 2.4 shows the reason why the first player can win in the first situation, but can only draw in the second situation.

\begin{figure}[H]
  \centering
  \includegraphics[width=8cm]{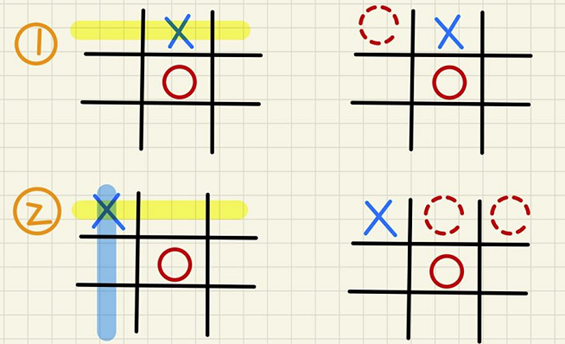}
  \caption{The differences between two situations}
\end{figure}

In Figure 2.4, if the opponent claims edge, it can only be used in one row (colored by yellow), so the strategy is designed to claim the corner next to the X first, “killing” the use value of X. On the contrary, if the opponent claims corner, it can be used in at least 2 rows (the yellow line and the blue line). The first player can’t simply “kill” this X through a single step now, but it is obvious that the first player should claim the colored position so that he can try his best to defend while attacking. But the problem is which position should he claim now? For example, supposing that the first player is going to claim the yellow line for the second step, should he claim the edge between X and O or claim the top right corner? Figure 2.5 shows the chain reactions in both cases.

\begin{figure}[H]
  \centering
  \includegraphics[width=6.5cm]{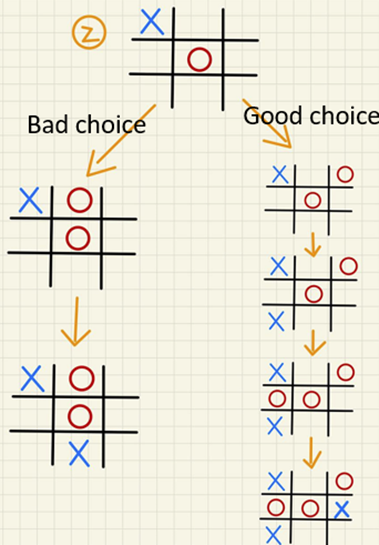}
  \caption{Chain reactions in both cases}
\end{figure}

The chain reaction was made by the normal prediction of both players. It is easy to predict that if one player has already claimed two thirds of a line, his opponent must claim the empty position of that line for the next move, otherwise he will lose the game. To handle and reduce the chaos in the chessboard, and better control the whole situation, the first should take the “Good choice” in Figure 2.5 obviously, which is a simple and fixed way for him to end up with a draw.

Unquestionably, if the opponent doesn’t perform as planned, which means he is not as smart as prediction, the first player should take the chance and win the game. An example is shown in Figure 2.6.

\begin{figure}[H]
  \centering
  \includegraphics[width=6cm]{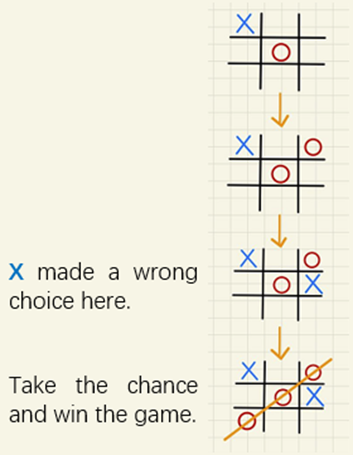}
  \caption{Take the chance and win}
\end{figure}

 \subsection{Misère version}
 \large{\bf Rule Introduction} In combinatorial games, the Normal Play Rule means that the last player to move wins, and the Misère Play Rule means the last player to move loses. Most of the conditions remain unchanged in misère version of tic-tac-toe, but it is declared that the first player to get three in a row loses, that is also why it is called “misère”.
 
 \begin{figure}[H]
  \centering
  \includegraphics[width=4.5cm]{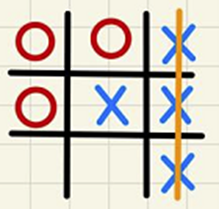}
  \caption{Red O wins in misère version}
\end{figure}

 \large{\bf Strategy Analysis} In misère version of tic-tac-toe, the second player should have a bit of advantage because he only takes 4 turns, while the first player takes 5 turns, which means the first player has more probability of making a line. Due to the fact that the first player takes disadvantage, his most appropriate target should be a draw.

First of all, it is very important for the first player to figure out where to begin with. There are three choices for the first player to start with: center, edge or corner, which has been shown in Figure 2.8. If the first player starts in the center, it belongs to 4 lines; if he occupies the edge, it belongs to 2 lines; if occupy the corner, it belongs to 3. To avoid making lines, should the first player avoid beginning in the center?  
 \begin{figure}[H]
  \centering
  \includegraphics[width=12cm]{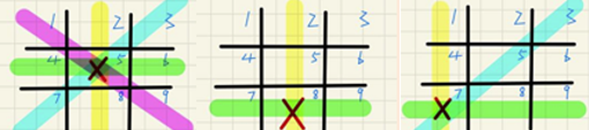}
  \caption{Different positions to start with}
\end{figure}

 However, the truth is exact opposite of what it looks like. The first player can always force a draw if he claims the center for the first move and then choose the square opposite of his opponent’s choice. The first player always loses if he claims edge or corner first, because there is always a winning strategy for his opponent in that case.
  \begin{figure}[H]
  \centering
  \includegraphics[width=8cm]{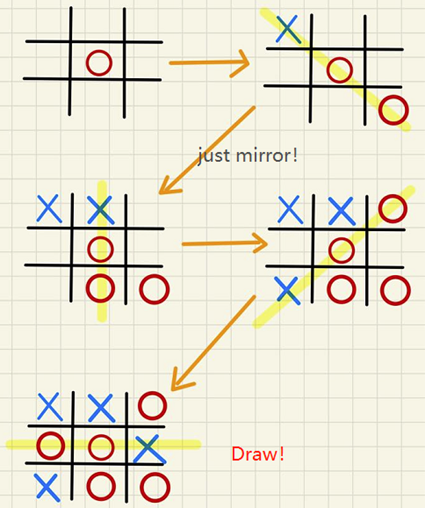}
  \caption{Mirror strategy in misère version}
\end{figure}
 Figure 2.9 shows how to force a draw for the first player by mirror his opponent. With claiming the center for the first move, and occupies the bottom right corner since his opponent occupied the top left corner. All of other steps are quite similar and simple. By the way, the mirror strategy mentioned here is a very useful method in many combinatorial games, like The Soda Can Game, Chomp and The Game of Nim.
   \begin{figure}[H]
  \centering
  \includegraphics[width=8cm]{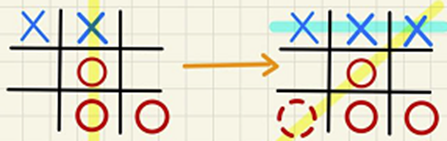}
  \caption{X loses first before O mirrors it}
\end{figure}
Here is an interesting case in Figure 2.10. If the second player knows the mirror strategy that the first player uses, he may try to “induce” the first player to get a three in a row, so he attempts to occupies the top right corner. But obviously, the second player would lose the game first because he got a 3-connection by himself! From this case, the first player should know there is no need to worry about anything, just mirror boldly as long as the game is not over.

To be emphasized, the only unbeatable way for the first player is described as above. If he starts with edge or corner, he always loses the game when facing a qualified opponent. 

There is the winning strategy for the second player if the first player makes a wrong choice. Assuming that the first player occupies the edge for the first move, the highlighted positions are labeled 1 in Figure 2.11, which means these positions have 1 amount of possibility to be used to make a row. From the perspective of the second player, he should try to force the first player to complete a row so he should avoid occupying the positions labeled 1. Instead, the second player could claim the position which contains the minimum number, like any position labeled 0 in Figure 2.11.

   \begin{figure}[H]
  \centering
  \includegraphics[width=8cm]{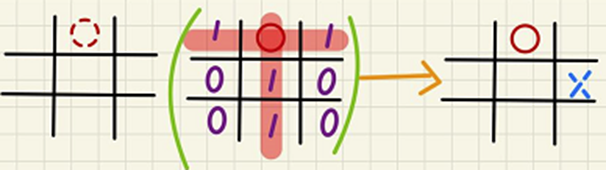}
  \caption{The second player analyzes the number and occupies a position labeled 0}
\end{figure}
In following steps, after the first player occupies any position, the second player should rebuild the “labeled graph” for every empty position. The logic is the same as before, if an empty position may be used to make a row by N chesses, it should be labeled N. After knowing all label numbers of every empty position, the second player can just take a position with the minimum number for his next move. But before the second player takes step, he should also check if that position would make himself lose, and he should change another position with the smallest number if the first position should not be occupied.

Figure 2.12 shows a whole process of the algorithm that the second player can use to force a win after the first player first claims edge. In the last few steps, when the situation is clear enough for the second player, he can almost claim any position and win without rebuilding the “labeled graph”. Of course, this strategy still works if the first player claims corner first.

   \begin{figure}[H]
  \centering
  \includegraphics[width=10cm]{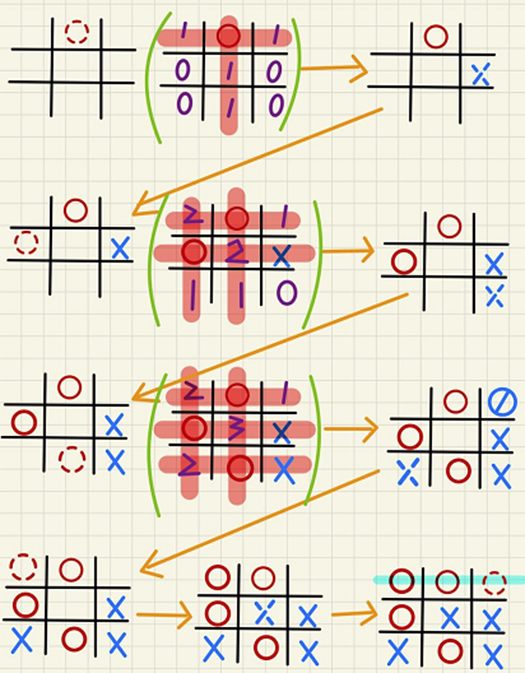}
  \caption{How the first player loses if he doesn’t claim center first in misère version}
\end{figure}

Actually, because the second player only takes 4 turns, so it is very easy for the second player to win even if he drops his chesses very casually. The algorithm mentioned above is just an auxiliary and feasible way for him to avoid making mistakes so that end up with a draw with such a big advantage. After all, it is a N-position for him, he should always win if he is a good player.

 \subsection{Reverse Misère version}
  \large{\bf Rule Introduction} The rule has changed again in reverse misère version of tic-tac-toe. All other rules remain the same, but both of players use X now, which means all the chesses in the board can be used together to complete a row. The player who makes a 3-connection loses, so it still belongs to misère play rule. Figure 2.13 shows a very simple case of losing result for red X, because red X got 3 in a row by its last step.

   \begin{figure}[H]
  \centering
  \includegraphics[width=4.5cm]{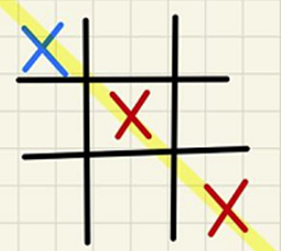}
  \caption{Red X loses in reverse misère version}
\end{figure}

 \large{\bf Strategy Analysis} In normal version and misère version of tic-tac-toe, there is always a possibility of draw, but in this game, the first player can always enjoy the N-position, which means he can always force a win if he moves first!

The winning strategy also begins in the center for the first player. As talk above, the first center belongs to 4 lines, which is exactly the 4 colors in Figure 2.14.

   \begin{figure}[H]
  \centering
  \includegraphics[width=8cm]{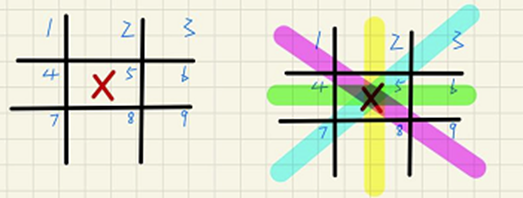}
  \caption{4 colors represent 4 lines crossing red X}
\end{figure}
In the following steps, every player’s move would occupy a position of colored area, which also means the square opposite of the position can no longer be occupied too. For example, if position No.3 was claimed, then anyone who claims the position No.7 would get 3 in a row and loses. To summarize, all of the four colors can only be used once.

   \begin{figure}[H]
  \centering
  \includegraphics[width=8cm]{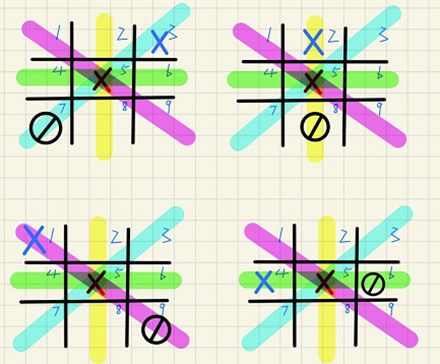}
  \caption{the square opposite of the occupied position can’t be claimed}
\end{figure}

Based on the conclusions above, it is easy to figure out the whole process and results for the two players. After the first player claims the center, the two players will take turns to occupy the colors. Because now it is the opponent’s turn, so the order should be: color1 for the second player, color2 for the first player, and color3 for the second player, color4 for the first player. Finally, the second player does not have new color to use so that he has to claim an occupied color and loses. That is the basic strategy for the first player to win in reverse misère version of tic-tac-toe.

The theory sounds very reasonable! But Figure 2.16 shows a practical result of the strategy, which contains something unexpected for the first player.

   \begin{figure}[H]
  \centering
  \includegraphics[width=10cm]{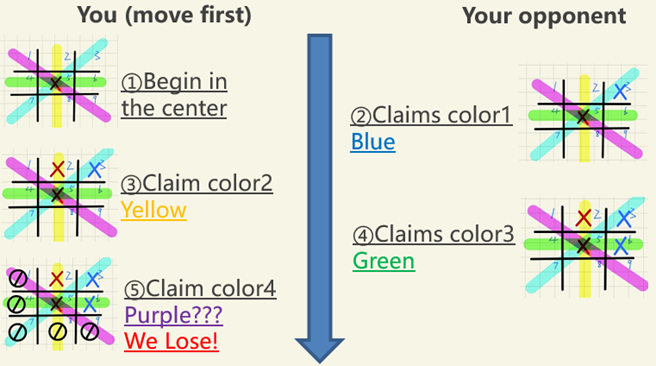}
  \caption{The first player can’t use the purple color and loses}
\end{figure}

It seems that something wrong with the basic strategy. The reason is the first player made a mistake that put his second chess too close to his opponent’s. So that it would be very easy for the chesses to be connected with each other. So, there should also be a strategy supplement, which is the first player should put his second chess far away from his opponent’s.

   \begin{figure}[H]
  \centering
  \includegraphics[width=8cm]{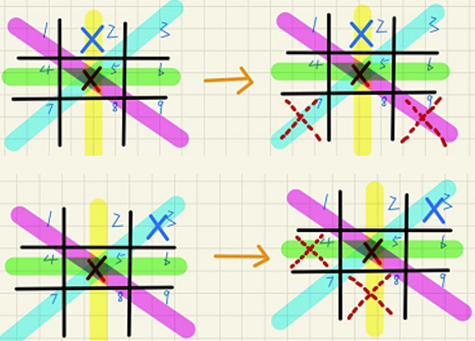}
  \caption{Strategy supplement}
\end{figure}

Figure 2.17 shows how the strategy supplement works. For example, if the second player first claims edge, like position No.2, the first player should claim position No.7 or position No.9; if the second player claims corner, like position No.3, the first player should claim position No.4 or position No.8. Then in the following steps, the first player can claim everywhere he wants in remaining colors, just remember don’t make a 3-connection by himself. And now he will definitely win.

   \begin{figure}[H]
  \centering
  \includegraphics[width=7cm]{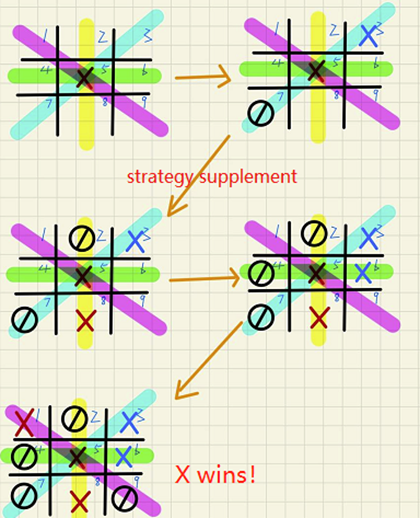}
  \caption{How the first player forces a win by final strategy in reverse misère version}
\end{figure}

An example of winning results for the first player (red X) is shown in Figure 2.18. The most key step the first player took is putting his second chess far away from his opponent. Unlike previous situation, now the first player can claim the purple line successfully after both of the two players took turns to occupy blue, yellow and green lines. Finally, all the empty positions are labeled “prohibition sign” so that there is no way to go for blue X.

\section{Algorithm design and implements}
This chapter will describe the program implementation of the strategies mentioned above in different kinds of tic-tac-toe. First it will come to the background which is all the same for every version, such as the drawings, interactions and so on. Then, for each kind of tic-tac-toe’s implementation, there will be descriptions of some basic modules, including the flowchart of algorithm in program, and there will also be display of result in the end. There will not be any specific code in this chapter, just a brief introduction about some simple but necessary parts.
 
\subsection{Basic module and Implementation}
The programs are written by Python, there is a useful package named Pygame, which is a free and open-source cross-platform library for the development of multimedia applications like video games using Python. Pygame can enable program to generate a beautiful interactive interface and accept user’s click actions. The programs used Pygame to draw the chessboard, accept the response of the player like when and where he puts his chess on and show the final result.

In general, the whole chessboard is a 700×700 pixel 2D coordinate plane, and every position is drawn with every single square. As for chesses like O and X, the former is just a circular and can be drawn easily by exact center position and radius; the latter can also be drawn through two straight lines perpendicular to each other, Figure 3.1 shows some of the examples. 

   \begin{figure}[H]
  \centering
  \includegraphics[width=7cm]{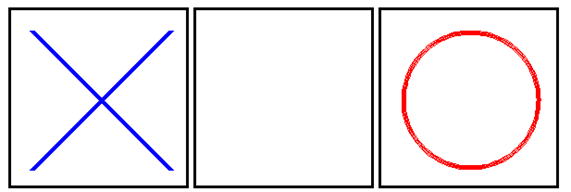}
  \caption{Some necessary elements in interface}
\end{figure}

The only key point to note is to clarify the conversion relationship between the "fake chessboard" on display and the "real chessboard" in memory, which is just a matrix size 3*3. For example, if one player clicks a location in interface, there would always be a calculation and transformation about the coordinate point in program, so that it can update the matrix in memory.

\subsection{Strategy implementation in Normal Version}

 \large{\bf Emergency Module (priority)} In normal version of tic-tac-toe, every player needs to discover and handle some urgent issues before trying to use so called “strategy”, because these directly related to his win or loss of the game. There are mainly two aspects to the emergency: stopping opponent when the opponent already completed two thirds of a line, completing a line to win the game when he has two thirds of a line of himself.

 \large{\bf “Clever-choice” Module} This module has the last priority to be used. Sometimes there is not any emergency to handle, and the strategy has been executed, the first player can almost occupy everywhere he wants. However, referring to the “labeled graph” in strategy analysis of Misère version, the first player can also choose a better position with the largest label number, which means it has more possibility to be used in a line. 
Maybe this module can’t change the result compared with random choices, but as long as it helps the first player perform better, it deserves to exist. Coincidentally, after the second player claims edge, this module can be used as a part of strategy.

 \large{\bf Flowchart of Strategy in Program} The following corresponds to the unbeatable strategy of normal version talked above, which will be integrate into flowchart of algorithm. The first player will occupy the center for his first move, and occupy the second position flexibly according to the opponent’s action. In flowing steps, if there is need of emergency module, execute it; if not, just turn to the “Clever-choice” Module. The flowchart is shown in Figure 3.2.

   \begin{figure}[H]
  \centering
  \includegraphics[width=14cm]{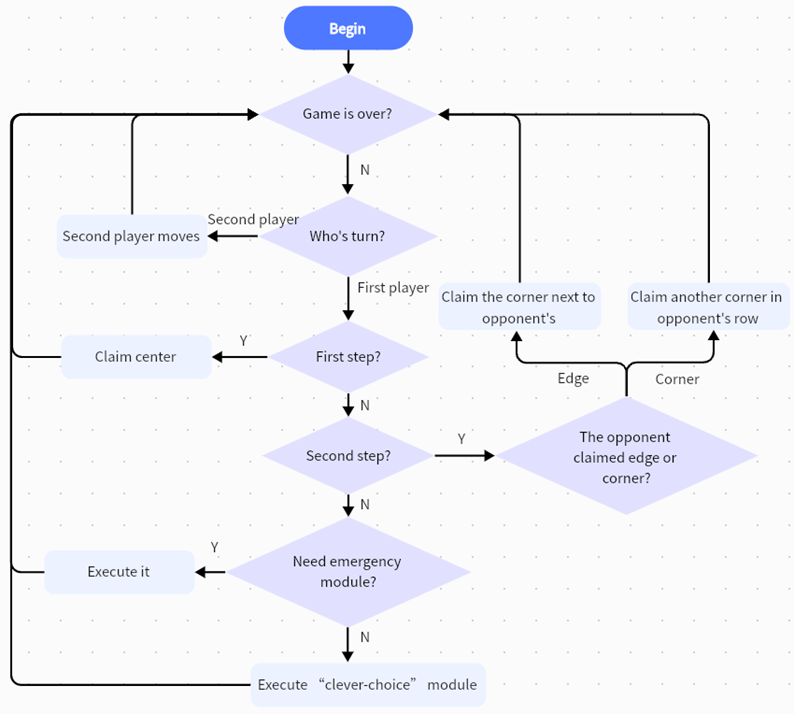}
  \caption{Flowchart of program in Normal Version}
\end{figure}

In figure 3.2, the program uses emergency module and clever-choice module to handle all the situations after the second step, which seems inconsistent with the unbeatable strategy in chapter 2.1, but they are essentially the same.

 \large{\bf Results Display} 
 
   \begin{figure}[H]
  \centering
  \includegraphics[width=12cm]{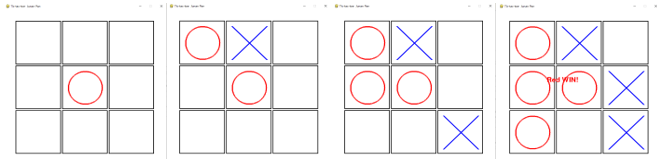}
  \caption{First player wins if the opponent claims edge first in Normal Version}
\end{figure}

   \begin{figure}[H]
  \centering
  \includegraphics[width=12cm]{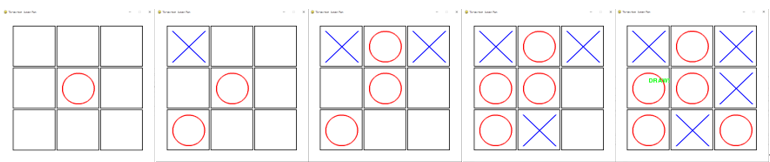}
  \caption{Draw if the opponent claims corner first in Normal Version}
\end{figure}

\subsection{Strategy implementation in Misère Version}

 \large{\bf Mirror Module} The mirror strategy in misère version of tic-tac-toe for the first player is easy to implement. After putting the first chess in the center, the program will catch the coordinate point of his opponent’s chess from Pygame interface so that it can calculate and claim the square opposite of his opponent.

 \large{\bf Flowchart of Strategy in Program} Figure 3.5 shows the flowchart of the strategy implemented in program. After the first step, the program always executes the mirror module, that is all the strategy.
 
    \begin{figure}[H]
  \centering
  \includegraphics[width=12cm]{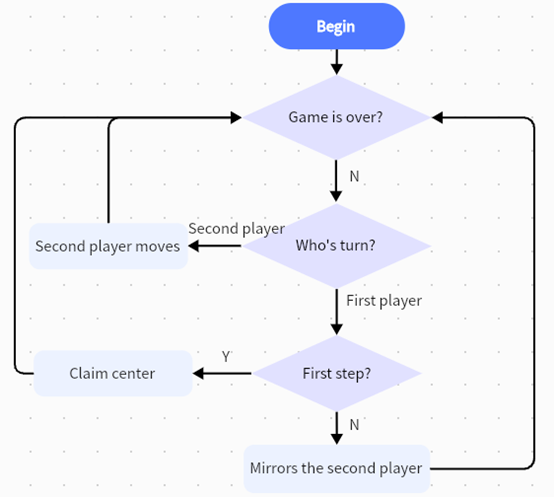}
  \caption{Flowchart of program in Misère Version}
\end{figure}

\large{\bf Results Display} 

    \begin{figure}[H]
  \centering
  \includegraphics[width=12cm]{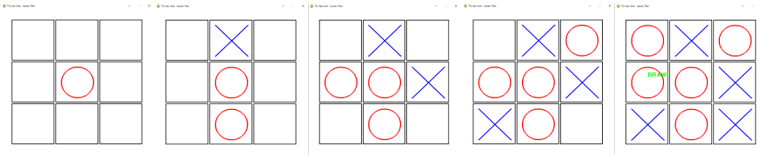}
  \caption{The first player force a draw by mirror strategy in Misère Version}
\end{figure}

In Figure 3.6, by always mirroring the second player, the program would never be defeated.

\subsection{Strategy implementation in Reverse Misère Version}
 \large{\bf “Keep distance” Module} This module comes from the conclusion of “mistake” that the first player may meet in game. It is clear that the first player should keep his second chess far away from his opponent. To keep the distance, the first player can claim the corner opposite of his opponent’s edge, or he can claim the edge opposite of the opponent’s corner. Given that the chessboard is a 3*3 matrix, the distance is always the same and fixed:$\sqrt{1^{2}+2^{2}}$, that is how the program finds the right place.

 \large{\bf Flowchart of Strategy in Program} In the strategy description of reverse misère version, 4 colors were used to show how the strategy works. But in implementation, there is no need for program to figure out the concept about colors. After claiming the center for the first move, the only strategy to do is “keep distance” from the opponent in the second step. Then in following steps, because the remaining available positions must belong to remaining available colors, so there is no longer any position better than others., they are all the same for every player. Figure 3.7 shows the flowchart, after two steps, the first player can claim any position in chessboard as long as it is safe.
 
     \begin{figure}[H]
  \centering
  \includegraphics[width=14cm]{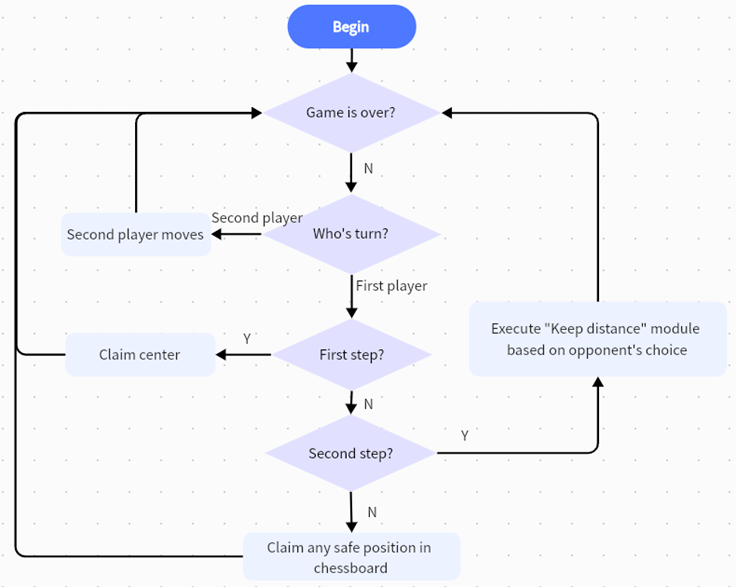}
  \caption{Flowchart of program in Reverse Misère Version}
\end{figure}

\large{\bf Results Display}

     \begin{figure}[H]
  \centering
  \includegraphics[width=12cm]{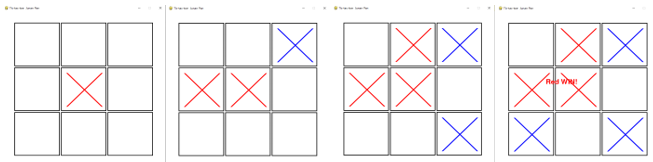}
  \caption{The first player can force a win by right strategy in Reverse Misère Version}
\end{figure}

% \section*{Acknowledgement}

\newpage
\section*{References}

[1] Crowley K, Siegler R S. Flexible strategy use in young children's tic-tac-toe[J]. Cognitive Science, 1993, 17(4): 531-561.

[2] Bolon T. How to never lose at Tic-Tac-Toe[M]. BookCountry, 2013.

[3] Beck J. Combinatorial games: tic-tac-toe theory[M]. Cambridge: Cambridge University Press, 2008.

[4] Hemanth D J. Analysis of Minimax Algorithm Using Tic-Tac-Toe[J]. 2020.

[5] Nasa R, Didwania R, Maji S, et al. Alpha-beta pruning in mini-max algorithm–an optimized approach for a connect-4 game[J]. Int. Res. J. Eng. Technol, 2018: 1637-1641.

[6] Larsson U, Nowakowski R J, Santos C P. Scoring games: the state of play[J]. Games of no chance, 2019, 5: 89-111.

[7] Nowakowski R J. Unsolved problems in combinatorial games[J]. Games of no chance, 2019, 5: 125-168.

[8] Hunt J. Building Games with pygame[M]//Advanced Guide to Python 3 Programming. Springer, Cham, 2019: 125-139.

% \section*{Thanks}

\end{spacing}   %这样写摘要两端不会缩进

%\end{multicols}

\end{document}